\documentclass[twocolumn]{aa}

\usepackage{graphicx}
\usepackage{txfonts}
\begin{document}
\title{On the mean field dynamo with Hall effect}

\author{A. Kandus, M.J. Vasconcelos,
\and A.H. Cerqueira}

\offprints{kandus@uesc.br}

\institute{
LATO-DCET-UESC, Rodovia Ilh\'eus-Itabuna km 16, Ilh\'eus, Bahia, 
45662-000, Brazil\\
\email{kandus@uesc.br, hoth@uesc.br, mjvasc@uesc.br}
}

\date{Received 29 August 2005 / Accepted 7 March 2006}

\abstract
{MHD turbulence with Hall effect.}
{Study how Hall effect modifies the quenching process of the electromotive
force (e.m.f.) in Mean Field Dynamo (MFD) theories.} {We write down the
evolution equations for the e.m.f. and for the large and small scale
magnetic helicity, treat Hall effect as a perturbation and integrate the
resulting equations, assuming boundary conditions such that the total
divergencies vanish.} {For force-free large scale magnetic fields,
Hall effect acts by coupling the small scale velocity and magnetic
fields. For the range of parameters considered, the overall effect is a
stronger quenching of the e.m.f. than in standard MHD and a damping of the
inverse cascade of magnetic helicity.} {In astrophysical environments
characterized by the parameters considered here, Hall effect would
produce an earlier quenching of the e.m.f. and consequently a weaker
large scale magnetic field.}

\keywords{Magnetohydrodynamics and plasmas -- Plasma turbulence}

\authorrunning{Kandus et al}
\maketitle

\section{Introduction}

The origin and evolution of magnetic fields observed in all objects of
the universe is one of the main problems in astrophysics. The basic
physical process assumed to create them is a dynamo, which needs two
basic ingredients, a seed field and an amplifying mechanism, each of them
constituting at present an independent line of research (e.g., Grasso \&
Rubinstein \cite{rev-dg}, Widrow \cite{rev-lw}, Giovannini \cite{rev-mg}).
An amplifying mechanism usually considered is the so called turbulent Mean
Field Dynamo (MFD), as turbulence is normally present in astrophysical
environments. In this mechanism it is assumed that turbulence is excited
at a small scale $\ell_s$ and that as a consequence a magnetic field
is induced at a larger scale $\ell_L$. This theory has been a useful
framework for modeling local origin of large scale magnetic fields in
stars and galaxies.

In the Universe there are very different astrophysical environments:
compact stars, low density and low temperature plasmas, accretion disks
around stars and in AGN's, etc. The plasma in each of those ambients has
a different composition and therefore different physical processes may
be relevant: in low ionized plasmas as the interstellar medium, ambipolar
difussion is important (Zweibel \cite{zweibel}); in the high-temperature
intracluster gas ohmic dissipation plays a major role, and Hall effect
can be relevant in accretion disks (Sano \& Stone \cite{stone}, Wardle
\cite{wardle99}, Balbus \& Terquem \cite{balb-terq}) as well as in
the early universe (Tajima et al. \cite{tajima}).  Turbulent dynamo
operation may therefore be affected by the composition of the plasma:
if we consider a plasma formed by, e.g.  protons, electrons and neutrals,
then the different interactions among these constituents can be expressed
as a generalized Ohm's law (Spitzer \cite{spitzer}, Priest \cite{priest}).

One of the main steps in the development of a turbulent mean field (or
large scale) dynamo theory was the recognition of the pivotal role played
by magnetic helicity (e.g., Pouquet et al. \cite{pfl}, Blackman \& Field
\cite{fb2000-1,fb2000-2}, Brandenburg \cite{bran-2001}). In the absence
of resistive dissipation, and for boundary conditions such that total
divergencies vanish, this quantity is globally conserved, independently of
any assumption about the turbulent state of the system. Its evolution does
not explicitly depend on the non-linear backreaction due to Lorentz force,
it merely depends on the induction equation, providing therefore a strong
constraint on the nonlinear evolution of the large scale magnetic field.

As stated above, mean field dynamo amounts to split the fields into large
scale mean fields ${\bf U}_0$, ${\bf B}_0$, ${\bf A}_0$ and small scale
turbulent fields ${\bf u}$, ${\bf b}$, ${\bf a}$. This small scale
fields represent the ``waste product'' of turbulence, and they can
be very intense in spite of their small coherence length \footnote{In
order to generate large scale fields, helical turbulence is needed. The
generation of these small scale fields is not to be considered as a result
of a {\sl small scale dynamo}.  Those dynamos require turbulence to be
non-helical (see e.g., Zel'dovich et al \cite{zeldovich}).}. In this
theory the evolution equation for ${\bf B}_0$ can be cast as $\partial
{\bf B}_0/\partial t = {\bf \nabla} \times \left({\bf U}_0\times {\bf
B}_0 + {\bf \varepsilon} - \eta {\bf J}_0\right)$, where ${\bf J}_0$ is
the mean electric current, $\eta$ the resisitivity and ${\bf \varepsilon}
= \langle {\bf u}\times {\bf b}\rangle_0$ the turbulent electromotive
force (e.m.f.).  In the two scale approach it is assumed that ${\bf
\varepsilon}$ can be expanded in powers of the gradients of ${\bf B}_0$ in
the rather general form $ \varepsilon_i = \alpha_{ij}\left(\hat {\bf g},
\hat{\bf \Omega},{\bf B}_0,...\right) B_{0j} + \eta_{ijk}\left(\hat{\bf
g},\hat {\bf \Omega},{\bf B}_0,...\right) \partial B_{0j} /\partial x_k$,
where the functions $\alpha_{ij}$ and $\eta_{ijk}$ are called turbulent
transport coefficients. They depend on the stratification $\hat {\bf g}$,
angular velocity $\hat {\bf \Omega }$ and mean magnetic field ${\bf
B}_0$.They may also depend on correlators involving the small scale
magnetic field in the form, for example, of small scale current helicity.

The simplest way of calculating the turbulent transport coefficients
consists of linearizing the equations for the small scale quantities,
ignoring quadratic terms that would lead to triple correlations in the
expressions for the quadratic terms. In other words, the backreaction
of mean field ${\bf B}_0$ on the correlation tensor of the turbulence
is taken into account, while neglecting the effect of the small
scale fields. In this way the electromotive force can be written as
(Krause \& R\"adler \cite{krause}) ${\bf \varepsilon} = \alpha {\bf
B}_0 - \beta {\bf J}_0$ with $\alpha \simeq -(1/3)\tau_{corr}\langle
{\bf u}\cdot {\bf \nabla}\times {\bf u}\rangle_0$ and $\beta \simeq
(1/3)\tau_{corr}\langle u^2\rangle_0$, with $\tau_{corr}$ the correlation
time.  These modifications of the turbulent transport coefficients have
been calculated about thirty years ago, and the approximation is known
as {\sl First order smoothing approximation}, or FOSA (see also Moffat
\cite{moffat2}, R\"udiger \cite{rudiger} Parker \cite{parker}; Moffat
\cite{moffat}; Zel'dovich, Ruzmaikin \& Sokoloff \cite{zeldovich}).

There remains to incorporate the modifications to ${\bf \varepsilon}$
that involve the small scale, fluctuating fields. They arise when
calculating $\langle {\bf u} \times {\bf b}\rangle_0$ from terms
involving the nonlinear terms and the Lorentz force in the evolution
equations for ${\bf b}$ and ${\bf u}$ respectively. As a consequence,
the $\alpha$ term written above gets renormalized in the nonlinear
regime by the addition of a term proportional to the current helicity
$\langle {\bf b}\cdot \left( {\bf \nabla}\times {\bf b}\right)$ of the
fluctuating field, which in turn is related to the magnetic helicity of
the small scale magnetic field. The $\beta$ term on the other side is not
affected by the backreaction of the small scale fields (Pouquet et all
\cite{pfl}, Subramanian \& Brandenburg \cite{sub-bran}, Brandemburg \&
Subramanian \cite{report-bs}). In this way we have $\alpha \simeq -(1/3)
\tau_{corr} \left( \langle {\bf u}\cdot \left({\bf \nabla} \times {\bf
u}\right)\rangle_0 - \langle {\bf b} \cdot \left( {\bf \nabla}\times
{\bf b}\right)\rangle_0\right)$.

In this paper we investigate how Hall effect modifies the process of
quenching of the e.m.f. ${\bf \varepsilon}$, described in the previous
paragraphs. For this purpose we use a closure scheme recently introduced
by Blackman \& Field (\cite{fb2002-2,fb2004}), that permits to study
dinamically the backreaction of both the large and small scale fields
on ${\bf \varepsilon}$. This closure, also named the ``minimal $\tau$
approximation'', consists in finding the evolution equation for the
electromotive force instead of finding ${\bf \varepsilon}$ itself.
In it, three point correlations of the generated small scale fields,
$\bar {\bf T}$, are not neglected, but their sum is assumed to be a
negative multiple of the second order correlator, i.e.  $\bar {\bf
T}=-{\bf \varepsilon}/\tau$. This assumption produces results that
are in very good agreement with numerical simulations (Brandenburg \&
Subramanian \cite{report-bs}).

Hall effect is taken into account by considering a corresponding term
in Ohm's law (see Spitzer \cite{spitzer}, Priest \cite{priest}).  The
parameter that meassures its intensity is the {\sl Hall length}, that in
alfv\'enic units is defined as $\ell_H = \left( 4\pi\rho\right)^{1/2}/n_e
e$, with $\rho$ the mass density and $n_e$ the electronic numerical
density. Of interest is the ratio of this length, to the Ohmic dissipation
length, $\ell_{\eta}$, and to the scale of the flow, $\ell_u$. For $\ell_H
\lesssim l_{\eta}$, the Hall-MHD equations reduce to those of standard
MHD, as ohmic dissipation erases any other interaction. In several
astrophysical problems, such as accretion disks, protoplanetary disks,
the early universe plasma and the magnetopause (Birn et al \cite{birn},
Balbus \& Terquem \cite{balb-terq}, Sano \& Stone \cite{stone}, Tajima
et al \cite{tajima}), the Hall scale is larger than the Ohmic scale,
but it can be smaller or larger than $\ell_u$.

Under the hypothesis that large scale fields are force-free (which is
a reasonable assumption in many astrophysical environments) we obtain
evolution equations for the mean magnetic field and for the large
and small scale magnetic helicities, that are formally identical to
those obtained in absence of Hall effect, provided that we redefine the
electromotive force by ${\bf \varepsilon}_H = \langle {\bf u}_e\times {\bf
b}\rangle$ with $u_e = u - \ell_H \langle \left( {\bf \nabla}\times {\bf
b}\right)\times {\bf b}\rangle_0$.  This means that the component of ${\bf
\varepsilon}_H$ along the mean field ${\bf B}_0$ governs the evolution of
magnetic helicity, and in turn magnetic helicity influences the growth
of ${\bf B}_0$ ( Parker \cite{parker}, Ji \cite{ji-99}).  We study a
system of coupled evolution equations for ${\bf \varepsilon}_H$ and
for the large and small scale magnetic helicities. As mentioned above,
the evolution equations for the helicities are formally identical to
the ones in absence of Hall effect. In contrast, the equation for ${\bf
\varepsilon}_H$ presents substantial differences in comparison to the
standard MHD case: {\sl (i)} In the $\alpha$ term, proportional to ${\bf
B}_0$, the fluid helicity is replaced by the electronic fluid helicity and
there appears an extra term, explicitly dependent on $\ell_H$ that couples
${\bf b}$ with ${\bf u}_e$.  {\sl (ii)} In the turbulent diffusion term,
proportional to ${\bf \nabla}\times {\bf B}_0$, the $\beta$ term, the
fluid kinetic energy term $\langle u^2\rangle_)$ is replaced by $\langle
{\bf u}\cdot {\bf u}_e\rangle_0$ and there also appears a correction
explicitly dependent on $\ell_H$, that couples ${\bf b}$ with ${\bf
u}$. All these modifications render this term not positive definite,
a fact that could produce a transfer of energy from small scales toward
large scales or, in other words an inverse cascade of energy (Mininni et
al. \cite{pablito4}).  {\sl (iii)} There appears a new term, proportional
to $\nabla^2 {\bf B}_0$, that again couples the mentioned velocities.
The coupling of ${\bf u}_e$ with ${\bf b}$ indicates that Hall effect
acts by transferring energy between these two fields in a non-trivial way.

In order to illustrate how Hall effect affects the quenching process of
the mean field dynamo, we applied the obtained equations to a specific
physical situation in which we considered that turbulence of maximal
kinetic helicity is excited at a certain scale $\ell_s$ and that the large
scale magnetic field is generated at a scale $\ell_L = 5\ell_s$. As for
the Hall effect, we considered $\ell_s < \ell_H <\ell_L$ and treated it
as a perturbation. We find that for this situation the overall effect
is a quenching of the e.m.f. stronger than in standard MHD, acompanied
by a  supression of magnetic helicity inverse cascade. Our results are
in qualitative agreement with recent numerical simulations performed by
Mininni et al (\cite{pablito3}).

As the aim of this paper is to understand conceptually how Hall effect
acts on a MFD, we did not apply our results to a concrete astrophysical
object. We leave this issue for future work, after a deeper understanding
of the mechanism is attained.  The paper is organized as follows:
in section \S2 we present the main equations and deduce the evolution
equations for large and small scale fields.  In section \S3 we deduce
the dynamo equations, i.e., the ones for the stochastic electromotive
force and for the large and small scale magnetic helicities. In section
\S4 we implement a two scale approximation and numerically integrate the
system of equations and discuss the results.  Finally in section \S5 we
sumarize our conclusions.

\section{Main Equations}

In Magnetohydrodynamics, Hall effect can be taken into account through
the generalized Ohm's law as (e.g., Spitzer \cite{spitzer}, Priest
\cite{priest}):

\begin{equation}
{\bf E}+{\bf U}\times {\bf B}=\frac{1}{n_e e}{\bf J}\times {\bf B}
+ \eta {\bf \nabla} \times {\bf B}\, ,
\label{a1}
\end{equation}

\noindent with ${\bf J}={\bf \nabla }\times {\bf B}$, $n_e$ the electron
number density, $e$ the modulus of the fundamental electric charge
and $\eta$ the Ohmic diffusion coefficient. We need the magnetic field
induction and the Navier Stokes equations.  We use units in which the
magnetic field has dimensions of velocity. To simplify the calculations
and the comparison with previous works we shall consider an incompressible
fluid, i.e., $\nabla \cdot {\bf U} = 0$\footnote{This condition implies
that the time that a sound signal takes to travel through a given distance
$l$ must be small compared to the time $\tau$ during which the flow
changes appreciably, i.e., $\tau \gg l/c_s$, so that the propagation
of interactions in the fluid may be regarded as instantaneous (e.g.,
Landau \& Lifshitz \cite{landau-fluids})}.  This condition is fulfilled
in several astrophysical environments. The Navier-Stokes equation is
then written as:

\begin{equation}
\frac{\partial {\bf U}}{\partial t} = -{\bf \bar P}\left[
\left( {\bf U}\cdot{\bf \nabla }\right) {\bf U}
- \left( {\bf B}\cdot {\bf \nabla}\right) {\bf B} \right]
- \nu {\bf \nabla}\times\left({\bf \nabla}\times {\bf U}\right)\, ,
\label{a2}
\end{equation}

\noindent where $ {\bf \bar P} \equiv {\bf I} - {{\bf \nabla} {\bf
\nabla}\cdot}/ {\nabla^2}$ is the projector operator onto the subspace of
solutions of this equation that satisfy the condition of incompressibility
(McComb \cite{macomb}) and $\nu$ the kinematic viscosity. The induction
equation reads:

\begin{equation}
\frac{\partial {\bf B}}{\partial t}={\bf \nabla }\times \left\{ {\bf U}
\times {\bf B}-\ell_H \left( {\bf \nabla }\times {\bf B}\right) \times
{\bf B}-\eta {\bf \nabla }\times {\bf B}\right\}\, , \label{a3}
\end{equation}

\noindent where we defined the {\it Hall length}, $\ell_H$, as $\ell_H =
{\left( 4\pi\rho\right)^{1/2}}/{n_e e}$. We also need the equation for
the vector potential ${\bf A}$. If we choose to work with the Coulomb
gauge, i.e., ${\bf \nabla}\cdot {\bf A}=0$, it reads:

\begin{equation}
\frac{\partial {\bf A}}{\partial t}= {\bf \bar P} \left[
{\bf U} \times {\bf B}-\ell_H \left({\bf \nabla }\times {\bf B}\right)
\times {\bf B}\right]
-\eta {\bf \nabla}\times {\bf B}\, , \label{a4}
\end{equation}

\noindent where ${\bf \bar P}$ is the previously defined projector, but
now projecting onto the space of functions that satisfy the chosen gauge.
When $\eta = 0$, Equation (\ref{a3}) represents the freezing of the
magnetic field to the electron flux. To see this, let us write ${\bf U}_e
= {\bf U} - \ell_H {\bf \nabla}\times{\bf B}$, which when substituted
in eqs (\ref{a3}) and (\ref{a4}) transforms them in equations formally
identical to the ones without Hall effect.

\subsection{Large and Small Scale Fields}

As we are interested in studying mean field dynamo, we split the fields
${\bf U}$, ${\bf B}$ and ${\bf A}$ as ${\bf U} =  {\bf u}$, ${\bf B} =
{\bf B}_0 + {\bf b}$ and ${\bf A} = {\bf A}_0 + {\bf a}$.  Upper case
and subindex $0$ denote large scale fields, i.e.  vector quantities
whose value may vary in space but whose direction and sense are almost
uniform or vary very smoothly. Technically speaking, they represent
\textit{ local spatial averages}.  Lowercase denotes small scale,
\textit{stochastic} fields, i.e. fields whose amplitude may be large,
but that have a very small coherence length. We assume that any average
of stochastic quantities is zero. Observe that we assumed ${\bf U}_0=0$,
i.e. no large scale flows.

\subsubsection{Evolution equation for the mean fields}

To derive the evolution equations for the large scale fields, we replace
the previous decomposition into eqs. (\ref{a3}) and (\ref{a4}) and take
\textit{local spatial averages} that we denote as $\langle ... \rangle_0$
\footnote{In the absence of large scale flows, i.e., if ${\bf U}_0 = 0$,
we obtain  from eq. (\ref{a2}) the following constraint: $\langle{\bf \bar
P}\left[ \left({\bf U}\cdot {\bf \nabla }\right) {\bf U} \right]\rangle_0
- \langle{\bf \bar P}\left[ \left({\bf b}\cdot {\bf \nabla }\right) {\bf
b}\right]\rangle_0 = 0$, that must be satisfied in order to guarantee
the vanishing of ${\bf U}_0$ for all time.}. If besides we demand large
scale fields to be force-free, we obtain:

\begin{equation}
\frac{\partial {\bf B}_0}{\partial t} =
{\bf \nabla }\times {\bf \varepsilon}_H +\eta \nabla^2 {\bf B}_0\, ,
\label{b1}
\end{equation}

\begin{equation}
\frac{\partial {\bf A}_0}{\partial t} =
{\bf \bar P} {\bf \varepsilon}_H +\eta \nabla^2 {\bf A}_0\, ,
\label{b2}
\end{equation}

\noindent where with the aid of Reynolds rules (McComb \cite{macomb})
we have interchanged derivatives with  averages. We have also defined
a {\it Hall turbulent electromotive force} as:

\begin{equation}
{\bf \varepsilon}_H = \langle {\bf u}\times {\bf b}\rangle_0
- \ell_H \langle \left( {\bf \nabla }\times {\bf b}\right)
\times{\bf b}\rangle_0 \equiv \langle {\bf u}_e \times {\bf b}\rangle_0\, .
\label{b3}
\end{equation}

\subsubsection{Evolution equations for the small scale fields}

The evolution equations for the small scale fields are obtained
by replacing the decomposition of fields into global averages and
stochastic component, into eqs. (\ref{a2}), (\ref{a3}) and (\ref{a4}) and
substracting from them the equations for the mean fields. Thus we have:

\begin{eqnarray}
\frac{\partial {\bf u}}{\partial t} &=&
{\bf \bar P}\left[ \left( {\bf b}\cdot{\bf \nabla }\right) {\bf B}_0
+ \left( {\bf B}_0\cdot{\bf \nabla }\right) {\bf b} \right]
+  {\bf \bar P}\left[ \left( {\bf b}\cdot{\bf \nabla }\right) {\bf b}
-  \left( {\bf u}\cdot{\bf \nabla } \right) {\bf u} \right] \nonumber\\
&-& \nu {\bf \nabla}\times\left({\bf \nabla}\times
{\bf u}\right)\, ,
~~~~~~~~~~~~~~~~~~~~~~~~~~~~~~~~~~~~~~~~~~~~~~~~~~~~~~~~~~
\label{b4}
\end{eqnarray}

\begin{eqnarray}
\frac{\partial {\bf b}}{\partial t} &=&
{\bf \nabla }\times \left( {\bf u}_e\times {\bf B}_0\right)
- \ell_H {\bf \nabla }\times \left[\left( {\bf \nabla }\times {\bf B}_0\right)
\times{\bf b}\right]
+ {\bf \nabla }\times \left( {\bf u}_e\times {\bf b}\right) \nonumber\\
&-& \langle {\bf \nabla }\times
\left( {\bf u}_e\times {\bf b}\right)\rangle_0
- \eta {\bf \nabla }\times \left({\bf \nabla }\times {\bf b}\right)\, ,
\label{b5}
\end{eqnarray}

\noindent and:

\begin{eqnarray}
\frac{\partial {\bf a}}{\partial t} &=&
{\bf \bar P} \left( {\bf u}_e \times {\bf B}_0\right)
- \ell_H {\bf \bar P} \left[\left({\bf \nabla }\times
{\bf B}_0\right) \times {\bf b}\right]
+ {\bf \bar P} \left( {\bf u}_e \times {\bf b}\right) \nonumber\\
&-& \langle{\bf \bar P} \left( {\bf u}_e \times {\bf b}\right) \rangle_0
-\eta {\bf \nabla}\times \left( {\bf \nabla}\times {\bf a}\right)\, .
\label{b6}
\end{eqnarray}

\section{Dynamo equations}

Now that we have the complete set of evolution equations for large and
small scale quantities, we can proceed to derive the evolution equations
for the electromotive force and the magnetic helicity.

\subsection{Magnetic helicity evolution equation}

Magnetic helicity is defined as the \textit{global average}, or
\textit{average over the entire volume} of ${\bf A}\cdot {\bf B}$,
that we denote by $H^M_T=\langle {\bf A}\cdot {\bf B}\rangle_{vol}$
(Biskamp \cite{biskamp}). These quantities do not vary in space, they
depend only on time.  Call $H^M \equiv \langle {\bf A}_0\cdot {\bf B}_0
\rangle_{vol}$ and $h^M \equiv \langle {\bf a}\cdot {\bf b}\rangle_{vol}$.
By taking the time derivative of these quantities with respect to time
and using eqs. (\ref{a3}), (\ref{a4}), (\ref{b5}) and (\ref{b6}) we obtain

\begin{eqnarray}
\frac{\partial H^M}{\partial t}
&=& 2\langle {\bf \varepsilon}_H \cdot {\bf B}_0 \rangle_{vol}
- 2 \eta \langle{\bf B}_0 \cdot\left({\bf \nabla} \times {\bf B}_0\right)
\rangle_{vol}\nonumber\\
&+& \langle {\bf \nabla}\cdot \left[ {\bf \varepsilon}_H \times {\bf A}_0
- \eta  \left( {\bf \nabla}\times {\bf B}_0 \right)
\times {\bf A}_0\right] \rangle_{vol}\, ,
\label{c1}
\end{eqnarray}

\noindent and:

\begin{eqnarray}
\frac{\partial h^M}{\partial t} 
&=& -2\langle {\bf \varepsilon}_H \cdot
{\bf B}_0\rangle_{vol} - 2\eta \langle \langle \left( {\bf \nabla}
\times {\bf b}\right) \cdot {\bf b}\rangle_0\rangle_{vol}\nonumber\\
&+& \langle {\bf \nabla} \cdot \left\{ \left[ {\bf u}_e \times  {\bf B}_0
-\ell_H \left( {\bf \nabla} \times {\bf B}_0\right) \times {\bf b}
\right] \times {\bf a}\right\}\rangle_{vol}\nonumber\\
&+& \langle {\bf \nabla} \cdot \left\{ \left[ {\bf u}_e \times {\bf b}
-\eta {\bf \nabla} \times {\bf b} \right] \times {\bf a}
\right\} \rangle_{vol}\, , \label{c2}
\end{eqnarray}

\noindent To deal with the operator ${\bf \bar P}$ we followed the
procedure deviced by Gruzinov \& Diamond (\cite{gruzinov-2}), that
consists in transforming Fourier the equations before taking averages,
and make a development to first order in $k_L/k_s$ with $k_L$ the scale
of $B_0$ and $k_s$ the scale of the small scale fields. We also made some
simple algebraic manipulation to put the total divergencies in evidence.
If we add up eqs (\ref{c1}) and (\ref{c2}) we see that total magnetic
helicity is conserved, except for the divergencies and the dissipative
terms. This means that the term $\langle {\bf \varepsilon}_H\cdot
{\bf B}_0\rangle_{vol}$ transforms magnetic helicity between mean and
fluctuating fields. In what follows we consider boundary conditions such
that the divergencies in eqs. (\ref{c1}) and (\ref{c2}) vanish. This
selection is debatable, however, in view of the fact that such conditions
may not be quite general, or easily attainable in practice. Nevertheless
they have two advantages: First, the resulting magnetic helicity is gauge
invariant and second, they are widely used in numerical simulations,
a fact that will facilitate comparisons with those works. The effect
of boundary conditions on the evolution and gauge invariance of
magnetic helicity is discussed in Berger \& Field (\cite{berger}), Ji
(\cite{ji-99}), Vishniac \& Cho (\cite{vish-cho}), and Subramanian \&
Brandenburg (\cite{sub-bran}).

\subsection{Evolution equation for ${\bf \varepsilon}_H^{\parallel}$}

According to its definition, eq. (\ref{b3}), ${\bf \varepsilon}_H$  is
the combination of two terms. So  we need to find evolution equations
for each term and then join them into one equation. The derivation is
sketched in Appendix {\bf A}, eq. (\ref{apa5b}), and here we quote the
final result, namely:


\begin{eqnarray}
\frac{\partial {\bf \varepsilon}_H}{\partial t}
&=&  \frac13 \left\{
- \langle {\bf u}_e \cdot \left( {\bf \nabla}\times {\bf u}_e\right) \rangle_0
+\langle \left( {\bf \nabla } \times {\bf b}\right)\cdot {\bf b}\rangle_0
+ \ell_H \langle {\bf b}\cdot \nabla^2 {\bf u}_e\rangle_0 \right\}{\bf B}_0\nonumber\\
&-&  \frac13 \left[\langle {\bf u}\cdot {\bf u}_e\rangle_0
+ \ell_H \langle {\bf b}\cdot \left( {\bf \nabla}\times {\bf u}\right) \rangle_0
 \right] \left( {\bf \nabla} \times {\bf B}_0 \right) \nonumber\\
&+&  \frac13\ell_H \langle {\bf u}_e\cdot {\bf b}\rangle_0 \nabla^2 {\bf B}_0
\nonumber\\
&+& \eta \left[\langle {\bf u}_e \times  \nabla^2 {\bf b} \rangle_0
 -\ell_H \langle \left( {\bf \nabla} \times \nabla^2
{\bf b} \right) \times {\bf b} \rangle_0 \right]\nonumber\\
&+& \nu \langle \nabla^2 {\bf u} \times {\bf b} \rangle_0 + {\bf \bar T}\, ,
\label{d20}
\end{eqnarray}

\noindent with ${\bf \bar T}$ representing the small scale
field, three-point correlations and given in Appendix {\bf A} by
eq. (\ref{apa6}).  The term proportional to ${\bf B}_0$ is similar to the
``$\alpha$'' term that appears in the kinematic dynamo, except that now
it has the \textit{electronic kinetic helicity} ($1^{st}$ term inside
braces) instead of the fluid kinetic helicity of the ordinary dynamo.
Besides this term, there is a current helicity term ($2^{nd}$ term
inside braces, also present in standard MHD dynamo equation) that is
due to the small scale magnetic field and that can be cast in terms of
the small scale magnetic helicity. Finally there is a new term that is
an explicit Hall modification ($3^{rd}$ term in braces), that couples
the small scale magnetic field to the electronic velocity field.

The term proportional to $\left( {\bf \nabla}\times {\bf B}_0\right)$,
named ``$\beta$ term'' is also strongly modified: the first term turns
out to be the scalar product of the fluid and kinetic velocities,
and there appears a second term, explicitly dependent on $\ell_H$ that
couples ${\bf u}$ to ${\bf b}$. All these modifications made this term
not positive definite anymore. A negative value of this coefficient
represents non-local transfer from small scale turbulent fields to the
large scale magnetic field (Mininni et al \cite{pablito4}). Finally
there appears a new term, proportional to $\nabla^2 {\bf B}_0$.

From eqs. (\ref{c1}) and (\ref{c2}) we see that the important quantity in
the mean field dynamo operation is the component of ${\bf \varepsilon}_H$
parallel to ${\bf B}_0$ (Parker \cite{parker}), which can be written
as ${\bf \varepsilon}_H^{\parallel} = {\bf \varepsilon}_H\cdot {\bf
B}_0/\vert {\bf B}_0\vert$ and whose evolution equation is then given by
${\partial {\bf \varepsilon}_H^{\parallel}}/{\partial t} = \left({\partial
{\bf \varepsilon}_H } /{\partial t} \right) \left( {{\bf B}_0}/{\vert
{\bf B}_0\vert}\right) + {\bf \varepsilon}_H \left[ \partial \left( {\bf
B}_0/\vert {\bf B}_0\vert\right) /{\partial t}\right]$. To numerically
integrate the resulting equation it is more convenient to write ${\bf
u}_e$ back in terms of ${\bf u}$ and ${\bf b}$. The physical reason
is that ${\bf u}$ is the velocity that can be externally excited or
prescribed. Thus we shall work with (see eq. \ref{apa5a} of Appendix.)

\begin{eqnarray}
\frac{\partial {\bf \varepsilon}_H^{\parallel}}{\partial t} &=&
\frac13 \left\{
\langle \left( {\bf \nabla } \times {\bf b}\right)\cdot {\bf b} \rangle_0
- \langle {\bf u} \cdot \left( {\bf \nabla}\times {\bf u}\right) \rangle_0
- \ell_H  \langle {\bf u} \cdot \nabla^2 {\bf b}\rangle_0  \right.\nonumber\\
&+& \ell_H \langle {\bf b}\cdot \nabla^2 {\bf u}\rangle_0
+ \ell_H \langle \left( {\bf \nabla}\times {\bf b}\right)
\cdot \left( {\bf \nabla}\times {\bf u}\right)\rangle_0 \nonumber\\
&-& \ell_H^2 \langle {\bf b} \cdot \left[ {\bf \nabla}\times
 \nabla^2 {\bf b}\right]\rangle_0
+\ell_H^2 \langle \left( {\bf \nabla} \times {\bf b}\right) \cdot
\nabla^2 {\bf b}\rangle_0 \left.\right\} \vert{\bf B}_0\vert \nonumber\\
&-&  \frac13 \left[\langle v^2\rangle_0 + \ell_H \langle  
{\bf \nabla} \cdot \left( {\bf u}\times {\bf b}\right)\rangle_0 \right]
\frac{\left( {\bf \nabla} \times {\bf B}_0 \right)\cdot {\bf B}_0}
{\vert {\bf B}_0\vert} \nonumber\\
&+& \frac13 \ell_H \left[ \langle {\bf u}\cdot {\bf b}\rangle_0
-\ell_H \langle \left( {\bf \nabla}\times {\bf b}\right) \cdot {\bf b}
\rangle_0 \right]
\frac{\left(\nabla^2 {\bf B}_0\right) \cdot {\bf B}_0}{\vert {\bf B}_0 \vert}
\nonumber\\
&-& \zeta_H^{\parallel} {\bf \varepsilon}_H^{\parallel}\, .
\label{e2}
\end{eqnarray}

\noindent where the last term represents the dissipative terms and
the more important the three-point correlations of the generated small
scale fields.

\section{Solving the System}

\subsection{Further approximations and numerical integration}

To numerically integrate the equations, we assume that full helical
turbulence is excited at a certain scale $\ell_S$ smaller than the
system's size, and that the large scale magnetic field is induced at
a larger scale $\ell_L = 5\ell_s$, that can be the system's size. This
assumption enables us to consider that spectra of small scale quantities
peak at wavenumber $k_S = 2\pi/\ell_S$ while large scale quantities do
so at $k_L = 2\pi/\ell_L$.  This assumption in based on the work of
Pouquet, Frish \& Leorat (\cite{pfl}), who several years ago showed
that when helical turbulence is induced at the scale $k_s$, large
scale quantities peaked at a smaller $k_L$ (see also Maron \& Blackman
\cite{maron}). Therefore we write the different terms of equations
(\ref{c1}), (\ref{c2}) and (\ref{e2}) as:
$\langle {\bf a.b}\rangle_{vol} = h^M_s$, 
$\langle {\bf A}_0.{\bf B}_0\rangle_{vol} = H^M_L$, 
$\langle {\bf u.b}\rangle_{vol} = h^C$, 
$\langle {\bf b}.\left({\bf \nabla} \times {\bf b} \right) \rangle_0 = k_S^2 h^M_s$, 
$\langle {\bf u}.\left( {\bf \nabla} \times {\bf u}\right) \rangle_0 =  h^u$,
$\langle {\bf u}. \nabla^2{\bf b}\rangle_0 = -k_S^2 h^C$, 
$\langle {\bf b}. \nabla^2{\bf u}\rangle_0 = -k_S^2 h^C$, 
$\langle \left( {\bf \nabla}\times {\bf b}\right) .  \left( {\bf \nabla}\times {\bf
u}\right) \rangle_0 = k_S^2 h^C$, 
$\langle \left( {\bf \nabla} \times {\bf b} \right) . \nabla^2 {\bf b}\rangle_0 = 
-k_S^4 h^M_s$, 
$\langle {\bf b}. \left( {\bf \nabla} \times \nabla^2 {\bf b}\right)
\rangle_0 = -k_S^4 h^M_s$, 
$\langle {\bf \nabla}\cdot \left( {\bf u}
\times {\bf b}\right) \rangle_0 = \pm k_S \vert \varepsilon_0\vert $,
$\nabla^2 {\bf B}_0 = -k_L^2 {\bf B}_0$. 
Besides we write $\langle {\bf u.u}\rangle_0 = 2 e^u$. We see that
in this case the the interaction between ${\bf u}$ and ${\bf b}$
is described by the small scale cross-helicity $\langle {\bf u}\cdot
{\bf b}\rangle_0 \equiv h^C$.  For ${\bf B}_0$, as it is force-fee,
we have $\vert {\bf B}_0\vert = k_L^{1/2} \vert h^M_L\vert^{1/2}$, and
$\left ({\bf \nabla}\times {\bf B}_)\right) . {\bf B}_0 = k_L^2 h^M_L$.
When we replace these expressions in eqs. (\ref{c1}) and (\ref{c2}) and
(\ref{e2}) we obtain:

\begin{eqnarray}
\frac{\partial \varepsilon_H^{\parallel}}{\partial t}
&=& \frac13 \left\{ k_S^2 h^M_s - h^u  + \ell_H k_S^2 h^C
\right\} k_L^{1/2}\vert h^M_L\vert^{1/2}\nonumber\\
&-&  \frac23 \left(e^u \pm \ell_H k_S  \vert \varepsilon_0\vert\right)
k_L^{3/2} \frac{h^M_L}{\vert h^M_L\vert^{1/2}}\nonumber\\
&-& \frac{\ell_H}3\left[ h^C - \ell_H k_s^2 h^M_s\right] k_L^{5/2}\vert 
h^M_L\vert^{1/2} - \zeta_H^{\parallel} {\bf \varepsilon}_H^{\parallel}\, , \label{f3}
\end{eqnarray}

\begin{equation}
\frac{\partial}{\partial t} h^M_L
= 2 k_L^{1/2}\varepsilon_H^{\parallel}\vert h^M_L\vert^{1/2}
- 2 \eta k_L^2h^M_L\,  \label{f6}
\end{equation}

\noindent
and

\begin{equation}
\frac{\partial}{\partial t} h^M_s
= - 2 k_L^{1/2}\varepsilon_H^{\parallel}\vert h^M_L\vert^{1/2}
- 2 \eta k_s^2h^M_s\,  \label{f7}
\end{equation}

\noindent $h^C$ is not an ideal invariant in Hall-MHD, as can be seen from
its evolution equation. It is obtained by deriving $\langle {\bf u}\cdot
{\bf b}\rangle_{vol}$ with respect to time, and using eqs. (\ref{b4})
and (\ref{b5}), and reads

\begin{eqnarray}
\frac{\partial \langle {\bf u}\cdot {\bf b}\rangle_{vol}}{\partial t} 
&=& 
- \ell_H \langle \left( {\bf \nabla}\times {\bf u} \right) 
\times \left( {\bf \nabla}\times {\bf b}\right) \rangle_0
 \cdot {\bf B}_0 \nonumber\\
&-& \ell_H \langle {\bf b} \times \left( {\bf \nabla}\times {\bf u}\right)
\rangle_0
 \cdot \left( {\bf \nabla}\times {\bf B}_0 \right) \nonumber\\
&+& \left( \nu + \eta \right) \langle \left( {\bf \nabla}\times {\bf u}\right)\cdot
\left( {\bf \nabla}\times {\bf b}\right)\rangle_{vol}\, . \label{f4}
\end{eqnarray}

In order to close our equation system, we could try to make in eq.
(\ref{f4}) the same approximations used in eq. (\ref{c1}), (\ref{c2})
and (\ref{e2}).  However they would produce expressions for which
new equations should be deduced.  These new equations in turn would
produce new terms and so on, thus resulting in a system difficult to
integrate and hard to interpret physically. Therefore we shall proceed as
follows. The presence of Hall effect implies that the magnetic field must
satisfy ${\bf \nabla}\times {\bf b}\not\propto {\bf b}$, i.e. it cannot
be force-free. Assuming ${\bf \nabla}\times {\bf b} \propto {\bf b}$
means two things: on one side that we are in the standard case (i.e.,
without Hall effect), and on the other, that the small scale magnetic
field is in an equilibrium state (i.e., no Lorentz force is excerted on
the stochastic electric currents). Therefore to use ${\bf \nabla}\times
{\bf b} \propto k_s {\bf b}$ in eq. (\ref{f4}) means to consider a
leading order in a perturbative expansion of the different terms of
eq. (\ref{f3}), around the Hall-free state, but it does not mean that
we are expanding around a force-free state, as the approximation is used
only in eq. (\ref{f4}). There remains the issue of the sign. As we shall
be interested in a situation in which small scale magnetic helicity grows
to negative values, we choose ${\bf \nabla}\times {\bf b} \simeq - k_s
{\bf b}$, to guarantee that condition. For the factor ${\bf \nabla}\times
{\bf v}$, we shall assume maximal negative helicity and thus write it
as ${\bf \nabla}\times {\bf v} \simeq - k_s {\bf v}$. We then write the
first and second terms in the r.h.s. of eq. (\ref{f4}) as $\langle \left(
{\bf \nabla}\times {\bf u} \right) \times \left( {\bf \nabla}\times {\bf
b}\right) \cdot {\bf B}_0 \rangle_{vol} \simeq  k_S^2 \langle {\bf u}
\times {\bf b} \rangle_{vol}\cdot {\bf B}_0$ and $\langle {\bf b} \times
\left( {\bf \nabla}\times {\bf u}\right) \cdot \left( {\bf \nabla} \times
{\bf B}_0 \right) \rangle_{vol} \simeq k_S \langle {\bf b} \times {\bf
u}\rangle_{vol}\cdot \left( {\bf \nabla}\times {\bf B}_0\right)$. Under
this approximation, we  note that the second term becomes smaller than
the first one by a factor $k_L/k_S$, and that therefore can be discarded
if $k_L/k_S \ll 1$. From the remaining expression, we see that we would
also need the evolution equation for ${\bf \varepsilon}_0\cdot {\bf
B}_0$. However as we are treating Hall effect as a perturbation, we make
a negligible error if we use ${\bf \varepsilon}_H\cdot {\bf B}_0$ instead
of ${\bf \varepsilon}_0\cdot {\bf B}_0$.  We use the same reasoning to
write $\vert \varepsilon_H\vert$ instead of $\vert \varepsilon_0\vert$
in the second term between brackets in the r.h.s. of eq. (\ref{f3}). We
are then left with the following equation for the cross helicity:

\begin{equation}
\frac{\partial h^C}{\partial t} =
-\ell_H k_S^2 \varepsilon_H^{\parallel} k_L^{1/2}\vert H^M\vert^{1/2}
- \left( \nu+\eta\right)k_s^2 h^C\, , \label{f5}
\end{equation}

\noindent and our equation system consists of eqs. (\ref{f3}), (\ref{f6}),
(\ref{f7}) and (\ref{f5}). In order to numerically integrate it and
to correctly devise the perturbative treatement of the Hall effect, we
need to make the equations nondimensional. We then define the following
dimensionless quantities:
$\tau = uk_L t$,
$G^M = h^M_L k_L/u^2$,
$g^M = h^M_s k_s/u^2$,
$g^C = h^C /u^2$,
$g^u = h^u/\left(k_Lu^2\right)$,
$Q_H^{\parallel} = \varepsilon_H^{\parallel} /u^2$,
$\lambda_H = \ell_H k_L$,
$\xi = \zeta / k_L u$,
$f^u = e^u/u^2$,
$R_M = u/\left( k_s\eta\right)$,
$r = k_S/k_L$.
This scheme of normalization is similar to the one of Blackman \& Field
(\cite{fb2002-2}), except that we use $k_L$ instead of $k_S$. Besides
we shall consider magnetic Prandtl number $\nu /\eta = 1$ and thus
$\left(\eta + \nu\right) = 2\eta$.  When we replace these quantities in
eqs. (\ref{f3}) and (\ref{f7})-(\ref{f5}) we obtain the following system:

\begin{eqnarray}
\frac{\partial Q_H^{\parallel}}{\partial \tau} 
&=& 
\frac13 \left[ g^M - \frac{g^u}{r^2} + \lambda_Hg^C\right] 
r^2 \vert G^M\vert ^{1/2} \nonumber\\
&-& \frac23 \left( f^u  \pm \lambda_H r \vert Q_H^{\parallel}\vert\right)
\frac{G^M}{\vert G^M\vert^{1/2}} \nonumber\\
&-&\lambda_H \left[ g^C -\lambda_H r^2 g^M\right]\vert G^M\vert^{1/2}
- \xi_H^{\parallel} Q_H^{\parallel}\, ,
\label{f8}
\end{eqnarray}

\begin{equation}
\frac{\partial G^M}{\partial \tau} = 2 Q_H^{\parallel} \vert G^M\vert^{1/2}
- \frac{2}{R_M r} G^M\, , \label{f9}
\end{equation}

\begin{equation}
\frac{\partial g^M}{\partial \tau} = - 2 Q_H^{\parallel} \vert G^M\vert^{1/2}
- \frac{2r}{R_M} g^M\, , \label{f10}
\end{equation}

\begin{equation}
\frac{\partial g^C}{\partial \tau} = -\lambda_H r^2 Q_H^{\parallel}
\vert G^M\vert^{1/2} - \frac{2r}{RM} g^C\, . \label{f11}
\end{equation}

\begin{figure}
\includegraphics[width=8.5cm]{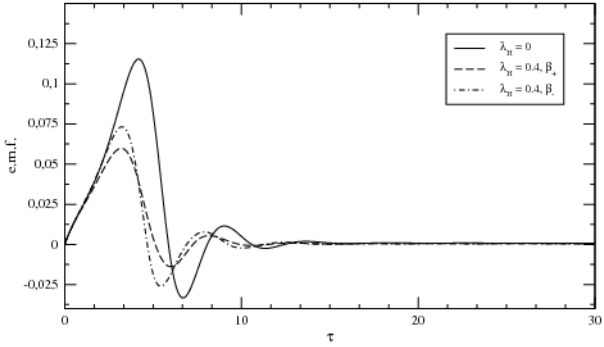}
\caption{Electromotive force for $\zeta = 1$, i.e. strong 
three-point correlations.}
\label{Figure1}
\end{figure}

In order to have a simple picture of how ${\bf \varepsilon}_H$ evolves,
we can reason as follows. For high $R_M$ assume a prescribed, negative
value for $g^u$ in the $\alpha$ term of eq. (\ref{f8}) (i.e, the first
term between square brackets), and that the term $\lambda_H g^C$ is
negligible as well as the initial value of $g^M$. In that situation  $G^M$
will initially grow toward positive values (due to the first term of
eq. [\ref{f9}]) and $g^M$ toward negative values (due to the first term
of eq. [\ref{f10}]). This will cause the $\alpha$ term to go to zero at
a certain instant, and hence to the end of the kinetic regime, i.e., the
period during which the growth of ${\bf B}_0$ is exponential. The presence
of the $\lambda_H g^C$ term drastically modifies this scenario: were this
term negative, then it would take the $g^M$ term a shorter time to cancel
the other two terms, i.e., we would have a shorter kinetic phase. Were
it positive, then the opposite situation would occur: the kinetic phase
would last longer\footnote{From the negative sign of the first term in
eq. (\ref{f11}) we see that in this case, the first situation will occur,
i.e. an earlier quenching of the dynamo}. This simple picture is even
more modified by the fact that now the $\beta$ term in eq. (\ref{f8})
(the second term between brackets) is not possitive definite, a fact that
could act in favour or against of the two situations described above. We
can conclude that the operation of a Hall-MHD dynamo is far more subtle
and complicated than the standard MHD one.

\subsection{Discussion}

To integrate the system we used a 4$^{th}$ order Runge-Kutta method
with variable step and considered the following values for the different
parameters that enter in the equations: $r=5$, $g^u=-5$ (this value is
equivalent to the $g^u=-1$ of Blackman \& Field \cite{fb2002-2} with the
normalization they used), $f^u = 1$, $R_M=2000$ and $\lambda_H = 0$ and
$0.4$. The second value of $\lambda_H$ corresponds to a Hall length almost
twice the turbulent scale, but shorter than the coherence large scale. The
high value of $R_M$ is easily found in astrophysical environments. For
the three-point correlations we considered two cases: $\zeta = 1$
(strong correlations) and $\zeta = 2/R_M$ (weak correlations). As initial
conditions we assumed $Q_{H0}^{\parallel} = 0$, $g^M_0 = 0.001 = G^M_0$
(i.e., an initial state with small magnetic helicity.  Other initial
conditions do not give qualitative different results) and $g^C=0$.
We also considered the two possible signs in the $\beta$ term, namely
$\beta_{\pm} = f^u \pm \lambda_H r \vert Q_H^{\parallel}\vert$ (see
eq. (\ref{f8})).

\begin{figure}
\includegraphics[width=8.5cm]{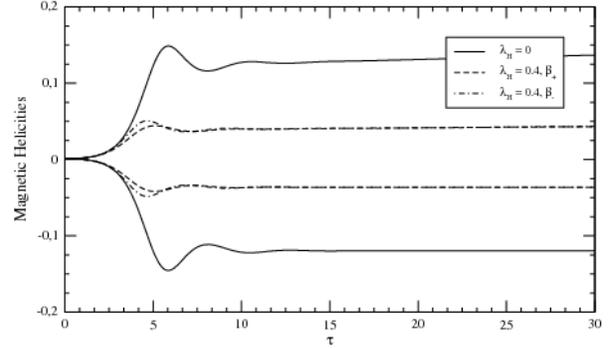}
\caption{
Magnetic helicities for $\zeta = 1$, i.e. strong three-point correlations. Upper
curves correspond to large scale MH, and lower curves to small scale MH. }
\label{Figure2}
\end{figure}

In Fig.\ref{Figure1} we plotted $Q_H^{\parallel}$ as a function of time,
for strong nonlinearities, i.e. $\zeta = 1$. We see that for $\lambda_H
\not= 0$, $Q_H^{\parallel}$ is damped faster than for $\lambda_H =
0$. For $\beta_+$ this process is in turn slightly stronger than for
$\beta_-$. The saturation value however is the same for all cases,
i.e. is not modified by Hall effect.  The rise of the first oscilation
in the transitory regime corresponds to the kinetic regime, in which
the large scale magnetic field would grow exponentially.  We see that
the instant at which this rise stops is slightly smaller than the one at
which stops the standard MHD curve, while the amplitudes of the curves
are substantially smaller. This instant is independet of either $\beta_+$
or $\beta_-$.  This behaviour can be interpreted as that the Hall dynamo
is less efficient than its standard MHD counterpart to generate large
scale fields.  In Fig. \ref{Figure2} we plotted small scale magnetic
helicity $g^M$ (lower curves) and large scale magnetic helicity $G^M$
(upper curves), also for $\zeta = 1$. We see that for $\lambda_H \not= 0$
the saturation value of $G^M$ is substantially smaller than for $\lambda_H
= 0$, and is the same for both possible $\beta$'s.  This means that the
inverse cascade of magnetic helicity is suppressed compared to standard
MHD, for the considered parameters.  Consistently with Fig. \ref{Figure1},
we see again that the rise of the first peak takes slightly less time
for $\lambda_H \not= 0$ than for $\lambda_H = 0$, and the amplitudes in
the former case are much smaller than in the latter case.

\begin{figure}
\includegraphics[width=8.5cm]{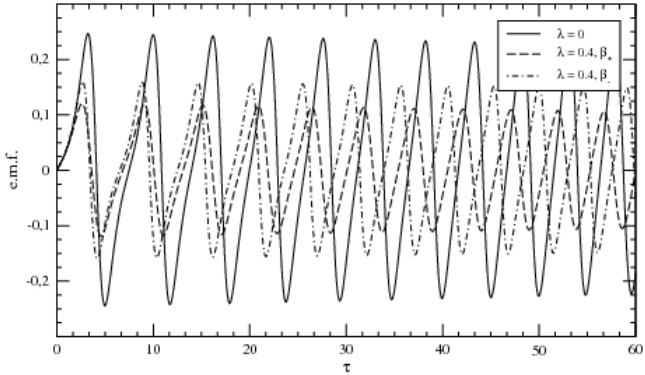}
\caption{Electromotive force for $\zeta = 2/RM$,
i.e. weak three-point correlations.}
\label{Figure3}
\end{figure}

\begin{figure}
\includegraphics[width=8.5cm]{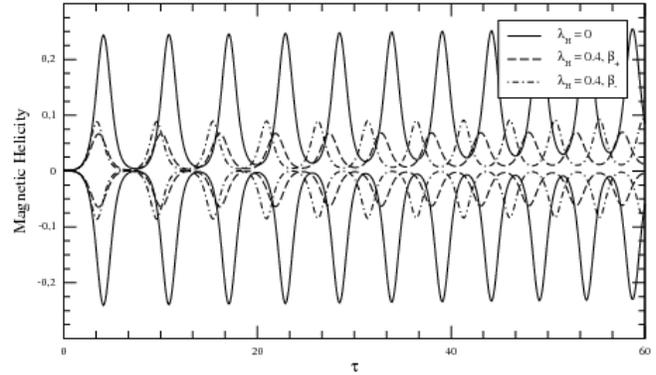}
\caption{Magnetic helicities for $\zeta = 2/RM$,
i.e. weak three-point correlations.}
\label{Figure4}
\end{figure}

In Fig. \ref{Figure3} we plotted the e.m.f. $Q_{H0}^{\parallel}$ for
$\zeta = 2/R_M$, i.e. weak non-linearities.  The smaller amplitudes of the
Hall-MHD curves means that the e.m.f. is more quenched than for standard
MHD, as in the case of strong non-linearities. Again in this case the
quenching due to $\beta_+$ is slightly stronger than the one produced
by $\beta_-$.  In this case the action of Hall effect in the e.m.f. is
manifested for all times, as no saturation value is attained. Again
here the rise of the first peak corresponds to the kinetic regime,
and similar features as for $\zeta=1$ are found: durantion slightly
shorter and amplitude significantly smaller, showing that in this case
again the Hall-dynamo is less efficient than the standard MHD one. In
Fig. \ref{Figure4} we plotted the small scale magnetic helicity $g^M$
(lower curves) and the large scale magnetic helicity $G^M$ (upper
curves). We can appreciate more clearly the quenching produced by Hall
effect, as the amplitude of the Hall dynamo is about 5 times shorter
than the standard MHD. The comments about the durantion of the kinetic
phase are the same as for the $\zeta=1$ case.

In all figures, the main features (oscillations for the $\zeta = 2/R_M$
and saturation for $\zeta = 1$) are determined by the $\alpha$ term,
i.e. by the interplay between kinetic and current small scale helicites,
and for the Hall dynamo also by the coupling between small scale magnetic
and velocity fields: Given a prescribed negative value of $g^u$ in the
$\alpha$ term of eq. (\ref{f8}), then $g^M$ grows negative, thus leading
to a cancellation of them and to a consequent supression of the growth
of $Q_{H0}^{\parallel}$ (this is the end of the kinetic regime). In the
Hall-dynamo, in the approximation we work with, the $\alpha$ term gets an
extra term proportional to $g^C$, which in principle can be positive or
negative. For the parameters we considered in this work it is negative and
consequently reinforces the action of $g^M$, thus leading to suppression
of the $\alpha$ term faster than in the non-Hall case. Were it positive,
then the opposite would occur: it would reinforce $-g^u$ and thus it
would take longer for the (negative) small scale magnetic helicity to
catch up with the other (positive) terms in the $\alpha$ term.  As a
result the kinetic regime would last longer.

The results quoted in this paper, namely a stronger quenching of the
Hall-MHD electromotive force for $\ell_{turbulence} < \ell_{Hall} <
\ell_{system}$, agree with numerical simulations performed by Mininni
et al (Mininni, Gomez \& Mahajan \cite{pablito3}).

\section{Conclusions}

In this paper we studied semianalitically how Hall effect modifies the
quenching process of the electromotive force in Mean Field Dynamo
theory.We used a dynamical closure scheme named minimal $\tau$
approximation (Blackman \& Field \cite{fb2002-2}, Brandenburg \&
Subramanian \cite{report-bs}), that takes into account the back-reaction
of the small scale fields generated by the turbulence. As we considered
helical turbulence, those small scale fields are not to be considered
as the result of a small scale dynamo, but as a waste product, that
nevertheless strongly quenches the e.m.f.

Considering force-free large scale magnetic fields, we found that Hall
effect modifies the evolution equation of the e.m.f. in several ways: the
main driving term, the so-called $\alpha$ term, proportional to $\vert
{\bf B}_0\vert$ now depends on the electronic velocity $u_e$ instead
of the fluid velocity. Besides there appears a third term, explicitly
dependent on the Hall parameter $\ell_H$, that couples the magnetic field
to this electronic velocity. The diffusive term, also known as $\beta$
term, proportional to $\vert {\bf \nabla}\times {\bf B}_0\vert$ also
depends on the electronic velocity but besides it acquired a new term
that couples the small scale magnetic field to the electronic velocity,
and that renders it not positive definite. A negative value of this
coefficient represents non local transfer from small scale turbulent
fields to the large scale magnetic field.  Finally there appears a new
term proportional to $\nabla^2 {\bf B}_0$. In our case this term plays
no significant role because it is substantially smaller than the others
due to the perturbative scheme we use.

To give a concrete numerical example, we considered Hall effect as a
perturbation of characteristic scale larger than the turbulent scale,
but shorter than the large scale magnetic field. This situation can be
found in e.g. accretion disks and in the early universe plasma (Sano \&
Stone \cite{stone}, Tajima et al. \cite{tajima}).  After implementing
the two scale approximation, we numerically integrated the resulting
evolution equations for the e.m.f., the large and small scale magnetic
helicities, and the cross-helicity which is the quantity that in this
approximation mimics the coupling between the small scale velocity
and magnetic fields. The overall effect is that in the presence of Hall
effect, the e.m.f. is more strongly quenched than in the case of standard
MHD dynamo. This fact is acompanied by a damping in the inverse cascade
of magnetic helicity.

As the main scope of this paper is to understand conceptually how
Hall effect acts on a MFD, we did not apply our results to a concrete
astrophysical object. We leave this issue for future work, after a
deeper understanding of the mechanism is attained.  Besides this point,
this work can be improved in several aspects. A very important one
is to extend the perturbative expansion to higher orders (or even to
device a non-perturbative scheme) in order to attain other regimes,
where dynamo action might be enhanced by Hall effect. This extension
might also mean that we should abandon the hypothesis that ${\bf B}_0$
is force-free. Next, to work with more general boundary conditions, that
permit to relax the conservation of total magnetic helicity. Finally,
a more detailed study of the three-point correlations is in order, as
well as to consider non-helical turbulence. We are working on some of
these topics at present.

\begin{acknowledgements}

We are grateful to E. G. Blackman who kindly and patiently clarified
several conceptual and technical aspects of his work and of dynamo
theory, and carefully read this manuscript. We also thank D. Gomez for
carefully reading and commenting this manuscript. A.K. acknowledges
financial support from FAPESB under grant APR0125/2005. M.J.V. thanks
PRODOC/UFBA (project n. 108). The work of AHC was partially supported
by a post-doctoral fellowship from the Brazilian agency CAPES (process
BEX 0285/05-6). AHC and MJV would like also to acknowledge A. Raga and
P. Vel\'azquez (ICN-UNAM), for their kind and warm hospitality during
our visit to the M\'exico City, where part of this work was developed.

\end{acknowledgements}

\begin{appendix}

\section{Evolution Equation for the Electromotive Force}

The evolution equation for $\varepsilon_H^{\parallel}$ is obtained by 
calculating

\begin{eqnarray}
\frac{\partial {\bf \varepsilon}_H}{\partial t} &=&
\langle \frac{\partial {\bf u}}{\partial t}\times {\bf b}\rangle_0
+ \langle {\bf u}\times\frac{\partial {\bf b}}{\partial t} \rangle_0 \nonumber\\
&-& \ell_H \left[\langle \frac{\partial \left( {\bf \nabla} \times {\bf b}\right) }{\partial t}
\times {\bf b}\rangle_0 +\langle \left( {\bf \nabla} \times {\bf b}\right)
\times \frac{\partial {\bf b}}{\partial t} \rangle_0\right]\, ,
\label{apa1}
\end{eqnarray}

\noindent and the corresponding equation for ${\bf
\varepsilon}_H^{\parallel} = {\bf \varepsilon}_H\cdot \left({\bf
B}_0/\vert {\bf B}_0\vert \right)$ by doing ${\partial {\bf
\varepsilon}_H^{\parallel}}/{\partial t} = \left({\partial {\bf
\varepsilon}_H } /{\partial t} \right) \left( {{\bf B}_0}/{\vert {\bf
B}_0\vert}\right) + {\bf \varepsilon}_H \left[ \partial \left( {\bf
B}_0/\vert {\bf B}_0\vert\right) /{\partial t}\right]$.  Replacing
eqs. (\ref{b4}) and (\ref{b5}), considering the development of operator
${\bf \bar P}$ to first order in the terms linear in ${\bf B}_0$, as is
done in Gruzinov \& Diamond (\cite{gruzinov-2}), and Blackman \& Field
(\cite{fb2002-2}), and assuming homogeneous and isotropic turbulence,
we obtain:

\begin{eqnarray}
\frac{\partial {\bf \varepsilon}_H}{\partial t}
&=& 
\frac13 \left\{ 
\langle 
\left( {\bf \nabla } \times {\bf b} \right)
\cdot {\bf b} \rangle_0 - \langle {\bf u} \cdot 
\left( {\bf \nabla}\times {\bf u}\right)
\rangle_0  - \ell_H  \langle {\bf u} \cdot
\nabla^2 {\bf b}\rangle_0 
       \right. 
\nonumber\\ 
&+& 
\ell_H \langle {\bf b}\cdot \nabla^2 {\bf u}\rangle_0
+ \ell_H \langle 
\left( {\bf \nabla}\times {\bf b} \right)
\cdot 
\left( {\bf \nabla}\times {\bf u}\right)
\rangle_0
\nonumber\\
&-& 
\left. 
\ell_H^2 \langle {\bf b} \cdot 
\left[ {\bf \nabla}\times \nabla^2 {\bf b}\right]
\rangle_0 +\ell_H^2 \langle 
\left( {\bf \nabla} \times {\bf b} \right) 
\cdot \nabla^2 {\bf b}\rangle_0 
\right\} 
{\bf B}_0 \nonumber\\
&-& \frac13 
\left[
\langle v^2\rangle_0 +
\ell_H \langle  {\bf \nabla} \cdot 
\left( {\bf u}\times {\bf b} \right)
\rangle_0 
\right] 
\left( {\bf \nabla} \times {\bf B}_0 \right)
\nonumber\\
&+&  \frac13 
\left[ 
\langle {\bf u}\cdot {\bf b}\rangle_0 - \ell_H \langle 
\left({\bf \nabla}\times {\bf b}\right)
\cdot {\bf b}\rangle_0
\right] 
\nabla^2{\bf B}_0 \nonumber\\
&+& \eta 
\left[\langle {\bf u}_e \times  \nabla^2 {\bf b} \rangle_0
 -\ell_H \langle 
\left( {\bf \nabla} \times \nabla^2 {\bf b} \right) 
\times {\bf b} \rangle_0 
\right]
\nonumber\\
&+& \nu \langle \nabla^2 {\bf u} \times {\bf b} 
\rangle_0 + {\bf \bar T}\, ,
\label{apa5a}
\end{eqnarray}

\noindent
where by ${\bf \bar T}$ we denote the non linear terms, i.e.:

\begin{eqnarray}
{\bf \bar T} 
&=& 
\langle {\bf u} \times \left[ {\bf \nabla }\times
\left( {\bf u}\times {\bf b}\right) \right] \rangle_0
+
\langle {\bf \bar P}\left(\left[ {\bf u}\times \left({\bf \nabla }\times
{\bf u} \right) \right] \times {\bf b}\right) \rangle_0 
\nonumber\\
&+& 
\langle {\bf \bar P}\left( \left[ \left( {\bf \nabla } \times
{\bf b}\right) \times {\bf b}\right] \times {\bf b}\right) \rangle_0
- \ell_H \langle{\bf u} \times\left\{
{\bf \nabla }\times \left[\left( {\bf \nabla }\times
{\bf b}\right) \times{\bf b}\right]\right\}  \rangle_0\nonumber\\
&-& \ell_H \langle \left( {\bf \nabla} \times {\bf b}\right)\times
\left[ {\bf \nabla } \times \left( {\bf u}\times {\bf b}\right) \right] \rangle_0
- \ell_H \langle \left\{ {\bf \nabla} \times\left[ {\bf \nabla }\times
\left( {\bf u} \times {\bf b}\right) \right]\right\} \times {\bf b} \rangle_0
\nonumber\\
&+& \ell_H^2 \langle \left( {\bf \nabla} \times {\bf b}\right)\times \left\{
{\bf \nabla } \times \left[\left( {\bf \nabla }\times {\bf b}\right)
\times{\bf b}\right]\right\} \rangle_0
\nonumber\\
&+& \ell_H^2 \langle \left[ {\bf \nabla} \times \left\{ {\bf \nabla }\times
\left[\left( {\bf \nabla }\times {\bf b}\right) \times{\bf b}\right]
\right\} \right] \times {\bf b} \rangle_0\, . \label{apa6}
\end{eqnarray}

Recalling that ${\bf u}_e = {\bf u} -\ell_H {\bf \nabla}\times {\bf b}$,
we can write eq. (\ref{apa5a}) in a form that shows
explicitly that now it is the electronic flow the driver of the dynamo:

\begin{eqnarray}
\frac{\partial {\bf \varepsilon}_H}{\partial t}
&=&  
\frac13 \left\{ \langle \left( {\bf \nabla } \times {\bf b}\right)\cdot
{\bf b} \rangle_0 - \langle {\bf u}_e \cdot 
\left( {\bf \nabla}\times {\bf u}_e\right) \rangle_0
+ \ell_H \langle {\bf b}\cdot \nabla^2 {\bf u}_e\rangle_0 \right\}{\bf B}_0
\nonumber\\
&-&  \frac13 \left[\langle {\bf u}\cdot {\bf u}_e\rangle_0
+ \ell_H \langle {\bf b}\cdot \left( {\bf \nabla}\times {\bf u}\right) \rangle_0
 \right] \left( {\bf \nabla} \times {\bf B}_0 \right) \nonumber\\
&+&  \frac13 \langle {\bf u}_e\cdot {\bf b}\rangle_0  \nabla^2{\bf B}_0
+ \eta \left[\langle {\bf u}_e \times  \nabla^2 {\bf b} \rangle_0
 -\ell_H \langle \left( {\bf \nabla} \times \nabla^2
{\bf b} \right) \times {\bf b} \rangle_0 \right]\nonumber\\
&+&\nu \langle \nabla^2 {\bf u} \times {\bf b} \rangle_0 + {\bf \bar T}\, .
\label{apa5b}
\end{eqnarray}

\end{appendix}
\end{document}